\documentclass[a4paper,10pt]{article}
\usepackage{jcappub}
\usepackage{a4wide}
\usepackage{amsfonts,amssymb,amsmath}
\usepackage{graphicx}
\usepackage{hyperref}
\usepackage{color}
\usepackage{eimodels}
\usepackage{eivars}

\usepackage{xspace}
\usepackage{lscape}

\makeatletter
\gdef\@fpheader{}
\makeatother

\hypersetup{
    bookmarks=true,         
    unicode=false,          
    pdftoolbar=true,        
    pdfmenubar=true,        
    pdffitwindow=false,     
    pdfstartview={FitH},    
    pdftitle={My title},    
    pdfauthor={Author},     
    pdfsubject={Subject},   
    pdfcreator={Creator},   
    pdfproducer={Producer}, 
    pdfkeywords={keyword1} {key2} {key3}, 
    pdfnewwindow=true,      
    colorlinks=true,       
    linkcolor=red,          
    citecolor=cyan,        
    filecolor=magenta,      
    urlcolor=blue,           
    linktocpage=true
}

\newcommand{\ie}{i.e.\xspace}

\newlength{\wsingfig}
\newlength{\wdblefig}
\newlength{\wfull}
\newlength{\hfull}
\newcommand{\sss}[1]{{\scriptscriptstyle{#1}}}

\newcommand{\usssMC}{\sss{\mathrm{MC}}}

\newcommand{\umod}{\sss{\mathrm{mod}}}

\newcommand{\calL}{\mathcal{L}}

\newcommand{\calM}{\mathcal{M}}

\newcommand{\Refc}[1]{Ref.~\cite{#1}}
\newcommand{\Refcs}[1]{Refs.~\cite{#1}}

\newcommand{\Fig}[1]{Fig.~\ref{#1}}

\DeclareMathOperator{\sech}{sech}

\newcommand{\GeV}{\mathrm{GeV}}

\newcommand{\Mpc}{\mathrm{Mpc}}

\newcommand{\muK}{\mathrm{\mu K}}

\newcommand{\OmegaCDM}{\Omega_\udm}
\newcommand{\OmegaB}{\Omega_\ub}

\newcommand{\thetaMC}{\theta_{\usssMC}}

\newcommand{\Clt}{C_{\ell}^{\mathrm{th}}}

\newcommand{\Clmock}{C_{\ell}^{\mathrm{mock}}}

\newcommand{\CnoiseT}{C_{\mathrm{noise}}^\mathrm{T}}
\newcommand{\CnoiseE}{C_{\mathrm{noise}}^\mathrm{E}}
\newcommand{\CnoiseB}{C_{\mathrm{noise}}^\mathrm{B}}
\newcommand{\Cnoise}{C_{\mathrm{noise}}}

\newcommand{\tetas}{\theta_{\mathrm{s}}}
\newcommand{\tetai}{\theta_{\uinf}}
\newcommand{\tetar}{\theta_{\ureh}}

\newcommand{\Mp}{M_\usssPl}

\newcommand{\fsky}{f_{\mathrm{sky}}}
\newcommand{\thetafwhm}{\theta_{\mathrm{fwhm}}}

\newcommand{\CAMB}{\texttt{CAMB}\xspace}
\newcommand{\COSMOMC}{\texttt{COSMOMC}\xspace}
\newcommand{\GetDistPlots}{\texttt{GetDistPlots}\xspace}
\newcommand{\ASPIC}{\texttt{ASPIC}\xspace}

\newcommand{\MULTINEST}{\texttt{MultiNest}\xspace}

\newcommand{\EI}{\textit{Encyclop{\ae}dia Inflationaris}\xspace}

\newcommand{\kstar}{k_*}

\newcommand{\Pstar}{P_*}

\newcommand{\wrehbar}{\overline{w}_{\ureh}}

\newcommand{\epsstar}[1]{\epsilon_{#1}}

\newcommand{\Treh}{T_\ureh}

\newcommand{\Nmod}{N^{\umod}}

\newcommand{\dwiTFIVE}{\dwi_{25}}
\newcommand{\dwiRTFIVE}{\dwi_{25}^{\mathrm{reh}}}

\newcommand{\ufid}{\mathrm{fid}}
\newcommand{\esiFID}{\mathrm{ESI}_\ufid}
\newcommand{\dwiFID}{\dwi_{\ufid}}
\newcommand{\mhiFID}{\mhi_{\ufid}}
\newcommand{\hiFID}{\hi_{\ufid}}
\newcommand{\lfiFID}{\lfi_{\ufid}}

\title{How Well Can Future CMB Missions Constrain Cosmic Inflation?}

\author[a]{J\'er\^ome Martin,}
\author[b]{Christophe Ringeval}
\author[a]{and Vincent Vennin}

\affiliation[a]{Institut d'Astrophysique de Paris, UMR
7095-CNRS, Universit\'e Pierre et Marie Curie, 98bis boulevard Arago,
75014 Paris (France)}

\affiliation[b]{Centre for Cosmology, Particle Physics and Phenomenology,
  Institute of Mathematics and Physics, Louvain University, 2 Chemin
  du Cyclotron, 1348 Louvain-la-Neuve (Belgium)}

\emailAdd{jmartin@iap.fr}
\emailAdd{christophe.ringeval@uclouvain.be}
\emailAdd{vennin@iap.fr}

\date{\today}

\begin{document}

\abstract{We study how the next generation of Cosmic Microwave Background
  (CMB) measurement missions (such as EPIC, LiteBIRD, PRISM and COrE)
  will be able to constrain the inflationary landscape in the hardest
  to disambiguate situation in which inflation is simply described by
  single-field slow-roll scenarios. Considering the proposed PRISM and
  LiteBIRD satellite designs, we simulate mock data corresponding to
  five different fiducial models having values of the tensor-to-scalar
  ratio ranging from $10^{-1}$ down to $10^{-7}$. We then compute the
  Bayesian evidences and complexities of all {\EI} models in order to
  assess the constraining power of PRISM alone and LiteBIRD
  complemented with the Planck 2013 data. Within slow-roll inflation,
  both designs have comparable constraining power and can rule out
  about three quarters of the inflationary scenarios, compared to one
  third for Planck 2013 data alone. However, we also show that PRISM
  can constrain the scalar running and has the capability to detect a
  violation of slow roll at second order. Finally, our results suggest
  that describing an inflationary model by its potential shape only,
  without specifying a reheating temperature, will no longer be
  possible given the accuracy level reached by the future CMB
  missions.}


\arxivnumber{1407.4034}

\maketitle

\section{Introduction}
\label{sec:intro}

Primordial gravity waves could play a crucial role in our attempts to
learn about the early Universe. They are a generic prediction of
inflation and their amplitude, described by the tensor-to-scalar ratio
$r$, carries precious information about the energy scale of
inflation. The observational situation of $r$, however, is at the time
of writing, unclear. The Cosmic Microwave Background (CMB) Planck data
puts an upper bound on $r$ (the precise value of which depends on the
assumptions made on the power spectra and on the priors; in a minimal
set up, one obtains $r\lesssim 0.11$) while the BICEP2 measurement of
the $B$-mode angular power spectrum would imply
$r=0.16^{+0.06}_{-0.05}$~\cite{Ade:2014xna, Ade:2014gua}. This last
value is currently debated and needs to be confirmed in view of the
role played by polarized foregrounds~\cite{Flauger:2014qra,
  Mortonson:2014bja, Ade:2014gna}. In this respect, the next release
of the Planck data will be of crucial importance, in particular in
order to assess the compatibility of these two data sets.

Recently, different CMB missions have been proposed and, among other
results, are expected to provide unprecedented constraints on
$r$. This includes the Experimental Probe of Inflationary Cosmology
(EPIC)~\cite{Bock:2008ww}, the Lite satellite for the studies of
B-mode polarization and Inflation from cosmic background Radiation
Detection (LiteBIRD)~\cite{Matsumura:2013aja}, the Polarized Radiation
Imaging and Spectroscopy Mission (PRISM)~\cite{Andre:2013nfa} and the
Cosmic Origins Explorer (COrE)~\cite{core}. These missions have
different designs and goals but, roughly speaking, they would allow a
measurement of a tensor-to-scalar ratio down to $r\simeq 10^{-3}$
(without assuming delensing~\cite{Sigurdson:2005cp,
  Namikawa:2014yca}). It seems therefore worth studying to which
extent such a measurement could improve our knowledge of inflation
compared to what has already been established with the Planck
data~\cite{Ade:2013uln, Martin:2013nzq}. In particular, it is
interesting to consider ``the hardest to disambiguate situation'' in
which inflation is well described by minimal single-field slow-roll
models. In this situation, which is the one favored by Planck 2013,
there is no entropy perturbations in the CMB, non-Gaussianities remain
unobservable while the primordial power spectra are
featureless~\cite{Chen:2012ja}. We choose to focus on this case since
this is the most difficult and conservative situation one can
imagine. Indeed, if realized, we won't be able to use the observables
mentioned before to narrow down the inflationary landscape. In other
words, if non-vanilla properties are detected by future missions, it
could only improve our ability to test and constrain inflation and,
therefore, the results discussed here represent what can be done in
the ``worst case scenario''.

In quantitative terms, the performance of a model, given a data set,
can be measured by calculating its Bayesian
evidence~\cite{Trotta:2005ar, Trotta:2008qt, Martin:2010hh,
  Easther:2011yq}. For all the {\EI} models~\cite{Martin:2014vha},
which currently achieve the best compromise between quality of the fit
and simplicity of the theoretical description (since, as already
mentioned, we do not observe entropy mode and/or non-Gaussianities),
Bayes factors have been recently computed for the Planck data in
\Refcs{Martin:2013nzq, Martin:2013gra} and for the BICEP2 data in
\Refc{Martin:2014lra}. A way to discuss the constraining power of an
experiment in this context is to use the Jeffreys' scale and count
the number of models in the ``inconclusive'', ``weak evidence'',
``moderate evidence'' and ``strong evidence'' zones with respect to
the best model. For instance, from the Planck data, one finds $26\%$
of the models in the first category (corresponding to $17$ different
shapes of the potential), $21\%$ in the second, $17\%$ in the third,
and $34\%$ in the fourth and last one. These numbers can be further
improved in one uses the Bayesian complexity as another statistical
indicator~\cite{Kunz:2006mc}. Of course, the Jeffreys' scale is
indicative only, although it is usually considered that models
belonging to the strong evidence category can really be considered as
``ruled out'', but the way the inflationary scenarios are distributed
among the Jeffreys' categories for different experiments gives a fair
view of their constraining power.

\begin{table}
\begin{center}
\begin{tabular}{| l | c | c | c | c | c |}
\hline
Satellite & $\CnoiseT$ & $\CnoiseE$ & $\CnoiseB$ & $\thetafwhm$ & $\fsky$ \\ \hline
PRISM &  $5\times 10^{-7}\, \muK^2$ & $2\CnoiseT$ & $2\CnoiseT$ & $ 3.2'$ &$0.7$ \\ \hline
LiteBIRD & $7\times 10^{-7}\, \muK^2$ &
$2\CnoiseT$ & $2\CnoiseT$ & $38.5'$ & $0.7$ \\ \hline
\end{tabular}
\caption{The two idealized CMB missions considered in this paper. They
  match the optimal specifications of LiteBIRD and PRISM and should be
  representative of other similar proposed designs (see text).}
\label{tab:expdes}
\end{center}
\end{table}

In this work, we simulate mock data for five fiducial models and two
representative future CMB missions corresponding to the idealized
design of LiteBIRD and PRISM (see Table~\ref{tab:expdes}). Then, using
the pipeline described in \Refcs{Ringeval:2013lea, Martin:2013nzq,
  Martin:2013gra}, we compute the Bayesian evidences of all {\EI}
models and their distribution among the Jeffreys' categories for the
two missions quoted before. This is our main result and it is
displayed in \Fig{fig:histo}.

This article is organized as follows. In the next section,
Sec.~\ref{sec:mock}, we describe the method used while the results are
presented in Sec.~\ref{sec:result}, namely the percentage of models in
each Jeffreys' categories for LiteBIRD and PRISM. Finally, the
implications for inflation and future strategies are discussed in the
conclusion in Sec.~\ref{sec:conclusion}.

\section{Methodology}
\label{sec:mock}

The cosmological model that we consider is an inflationary flat
$\Lambda$CDM scenario. It is characterized by the parameters
$\theta=\{\tetas, \tetar, \tetai\}$ where the quantities $\tetas$,
given by $\tetas\equiv \{\OmegaB h^2, \OmegaCDM h^2, \tau,
100\thetaMC\}$ (respectively, the baryons normalized density, the cold
dark matter normalized density, the optical depth and an angle related
to the angular size of the sound horizon on the last scattering
surface; $h$ being the reduced Hubble parameter), describe
post-inflationary physics, the $\tetar$'s describe the reheating phase
and the $\tetai$'s are inflationary parameters describing the shape of
the inflaton potential~\cite{Martin:2010kz, Dai:2014jja}. The initial
power spectra for density perturbations and primordial gravity waves
are given by formulas derived in the slow-roll approximation, \ie a
scale-invariant piece plus small scale dependent logarithmic
corrections~\cite{Hoffman:2000ue, Schwarz:2001vv}. As a consequence,
they depend on the parameter $\Pstar$, describing the overall
normalization of the primordial fluctuations and three slow-roll
parameters (at second order in slow roll) evaluated at Hubble radius
crossing during inflation, $\epsilon_1$, $\epsilon_2$ and
$\epsilon_3$; for explicit formulas, see for instance
\Refcs{Martin:2013uma, Jimenez:2013xwa, Martin:2014vha}. The
dependence on specific inflationary scenarios lies in the fact that
$\epsilon_n=\epsilon_n\left(\tetai,\tetar\right)$.

\begin{table}
\begin{center}
\begin{tabular}{|l|c|c|}
\hline
Fiducial model & Potential $V(\phi)/M^4$ & Potential parameters  \\ \hline
$\lfiFID$ & $\left(\phi/\Mp\right)^2$ & \\ \hline
$\dwiFID$ & $\left[(\phi/\phizero)^2-1\right]^2$ & $\phizero/\Mp=25$  \\ \hline
$\hiFID$ & $\left[1-\exp\left(-\sqrt{2/3}\phi/\Mp\right)\right]^2$ & \\ \hline
$\esiFID$ & $1-\exp\left(-q\frac{\phi}{\Mp}\right)$ & $q=8$  \\ \hline
$\mhiFID$ & $1-\sech\left(\phi/\mu\right)$ & $\mu/\Mp=0.01$  \\ \hline
\end{tabular}
\caption{The five fiducial models used to generate the mock data and
  the corresponding potential parameters value. The
  post-inflationary cosmological parameters have been fixed to typical
  values compatible with the Planck 2013 data, namely $\OmegaB
  h^2\equiv0.0223$, $\OmegaCDM h^2\equiv0.120$, $\Omega_{\nu}
  h^2\equiv6.45\times 10^{-4}$, $\tau \equiv0.0931$ and $h \equiv
  0.674$. The reheating temperature has been set to $\Treh \equiv 10^8
  \, \GeV$ with a reheating mean equation of state $\wrehbar \equiv 0$
  and a primordial amplitude for the scalar perturbations $\Pstar
  \equiv 2.203\times 10^{-9}$.}
\label{tab:fidmod}
\end{center}
\end{table}

\begin{table}
\begin{center}
\begin{tabular}{| l | c | c | c | c | c |}
\hline
Fiducial model & $\epsilon_{1}$ & $\epsilon_{2}$ & $\epsilon_{3}$ & $\nS$ & $r$ \\ \hline
$\lfiFID$ & $9.63\times 10^{-3}$ &
$1.93\times 10^{-2}$ & $1.93\times 10^{-2}$ & $0.961$ & $1.52\times 10^{-1}$ \\ \hline
$\dwiFID$ &
$5.40\times 10^{-3}$ & $2.77\times 10^{-2}$ & $1.41\times 10^{-2}$ &
$0.962$ & $8.45 \times 10^{-2}$ \\ \hline
$\hiFID$  & $2.65\times 10^{-4}$ & $3.81\times 10^{-2}$ & $1.93\times 10^{-2}$ & $0.961$ & $4.12 \times 10^{-3}$ \\ \hline
$\esiFID$ & $3.28\times 10^{-6}$ & $4.10\times 10^{-2}$ & $2.05\times 10^{-2}$ & $0.959$ & $5.09 \times 10^{-5}$ \\ \hline
$\mhiFID$  & $2.19\times 10^{-8}$ & $4.19\times 10^{-2}$ & $2.09\times 10^{-2}$ & $0.958$ & $3.40 \times 10^{-7}$ \\ \hline
\end{tabular}
\caption{Fiducial values for the slow-roll parameters, the spectral index
  and the tensor-to-scalar ratio (at the pivot scale
  $\kstar=0.05\,\Mpc^{-1}$) for the five fiducial models used to
  generate the mock data (see Table.~\ref{tab:fidmod}).}
\label{tab:fidval}
\end{center}
\end{table}

Let us now present the line of reasoning used in this article. The
first step has been to generate mock data. We have chosen to analyze
five different situations, associated with five fiducial models
summarized in table~\ref{tab:fidmod}.

The first one corresponds to a case where B-modes should easily be
detected by the future experiments because the underlying model is
compatible with the BICEP2 value, $r\simeq 0.16$. To model this case,
we have chosen a fiducial slow-roll inflationary scenario given by the
quadratic Large Field model $m^2\phi^2/2$ ($\lfiTWO$ in the {\EI}
terminology). The second case is supposed to describe a situation
where the estimated value of the tensor-to-scalar ratio is smaller
than the BICEP2 value due to a possible re-estimation of the polarized
foregrounds contribution, say $r\simeq 0.08$~\cite{Flauger:2014qra,
  Mortonson:2014bja}. To describe this case, we have considered a
Double Well inflation scenario ($\dwi$) with $\phizero/\Mp \equiv
25$. The third situation corresponds to a case where one is close to
the detection limit of $r$ and we use the Starobinsky (or Higgs)
inflation model ($\hi$) to parametrize this situation. The fourth
example is chosen such that the tensor-to-scalar ratio is less than
the threshold value $r=10^{-3}$ and we consider an Exponential
Supersymmetric scenario ($\esi$) with $q \equiv 8$. Finally, the fifth
and last case corresponds to a situation where the fiducial model is
associated with an undetectable amount of primordial gravity waves. To
describe this possibility, we have chosen a Mutated Hilltop inflation
model ($\mhi$) with $\mu\equiv0.01\Mp$ (see \Refc{Martin:2014vha} for
more details on these models).

Moreover, all fiducial models share the same reheating and
cosmological parameters. The reheating temperature has been fixed to
$\Treh\equiv10^{8}\,\GeV$ with a mean equation of state
$\wrehbar\equiv0$ while the post-inflationary evolution is assumed to
be described by a flat $\Lambda$CDM model with $\OmegaB
h^2\equiv0.0223$, $\OmegaCDM h^2\equiv0.120$, $\Omega_{\nu}
h^2\equiv6.45\times 10^{-4}$, $\tau \equiv0.0931$ and $h \equiv
0.674$. For each fiducial model, the first three Hubble flow functions
have been calculated using the {\ASPIC}
library\footnote{\url{http://cp3.irmp.ucl.ac.be/~ringeval/aspic.html}}
and they are given in table~\ref{tab:fidval}. For each model, the
{\ASPIC} code solves the reheating consistent slow-roll equations to
get the field value at which the pivot scale $\kstar$ crossed the
Hubble radius during inflation, and then the corresponding values for
the Hubble flow functions (see section~2.2 in \Refc{Martin:2014vha}).
At last, we have used a modified version of the {\CAMB} code to
generate the fiducial temperature and polarization multipole
moments~\cite{Lewis:1999bs}.
  
Once these mock data have been generated, the next step is to analyze
them from the point of view of the two CMB missions under focus and
this requires to specify their likelihood function. For this purpose,
assuming Gaussian statistics, one can show that the likelihood
function $\calL\left[\Clmock \vert
  \Clt\left(\tetas,\tetar,\tetai\right),\Sigma\right]$ over the full
sky is given by a Wishart distribution~\cite{Bond:1998qg,
  Perotto:2006rj, Percival:2006ss, Hamimeche:2008ai}. Here $\Clmock$
and $\Clt$ stand for the mock and theoretical angular power spectra,
respectively. These distributions are different for the two
experiments under scrutiny, PRISM and LiteBIRD, due to their different
specifications and this is encoded in the quantity $\Sigma \equiv
\{\CnoiseT,\CnoiseE,\CnoiseB,\thetafwhm,\fsky\}$. The noise power
$\Cnoise$ for temperature and polarization, the full width at half
maximum (fwhm) $\thetafwhm$ for the (assumed) Gaussian beam and
fraction of sky coverage $\fsky$ are summarized in
Table~\ref{tab:expdes}. As can be seen in this table, the LiteBIRD
beam resolution is relatively poor, by design, and as such, we have
complemented the LiteBIRD forecast with the Planck 2013 data.

In order to carry out a Bayesian analysis, one must specify the
priors. Here, we have chosen the same priors on $\tetas$, $\tetar$ and
$\tetai$ than those discussed at length in \Refcs{Martin:2013nzq,
  Martin:2014lra}. Bayesian evidences and complexities are then
derived by performing a data analysis for each experiment and each
fiducial model by using all the model of {\EI}. Moreover, we have
assumed uninformative priors between the different models $\calM_i$,
that is to say $\pi(\calM_i)=1/\Nmod$, where $\Nmod$ is the number of
{\EI} scenarios that we consider. In practice, we have followed the
same method as described in \Refcs{Ringeval:2013lea, Martin:2013nzq,
  Martin:2014lra} which involves the derivation of an effective
marginalized likelihood function depending only on the reheating and
primordial parameters. All evidences have been derived using the
{\MULTINEST} nested sampling algorithm~\cite{Mukherjee:2005wg,
  Feroz:2007kg, Feroz:2008xx} with a target accuracy of $10^{-4}$ and
a number a live points equals to $30000$.

In the following section, we report the results that have been
obtained for all the {\EI} scenarios. To make the comparison with the
current Planck (and BICEP2) constraints realistic, let us stress that
the fiducial models, having zero free parameters and used to generate the
mock data, \emph{are not} included in the list of models tested.

\section{Results and Discussion}
\label{sec:result}

\begin{figure}
\begin{center}
\includegraphics[width=\wdblefig]{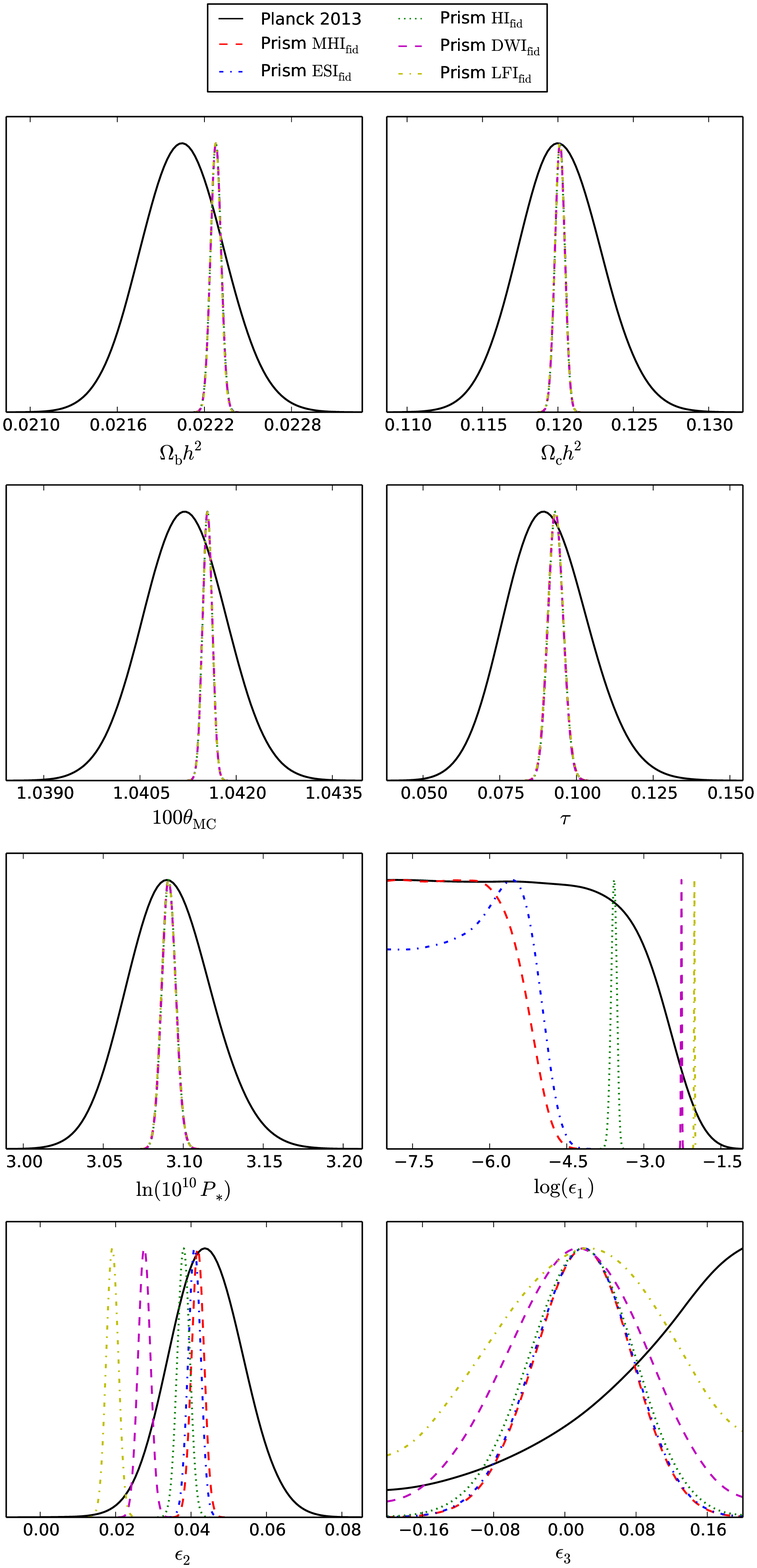}
\includegraphics[width=\wdblefig]{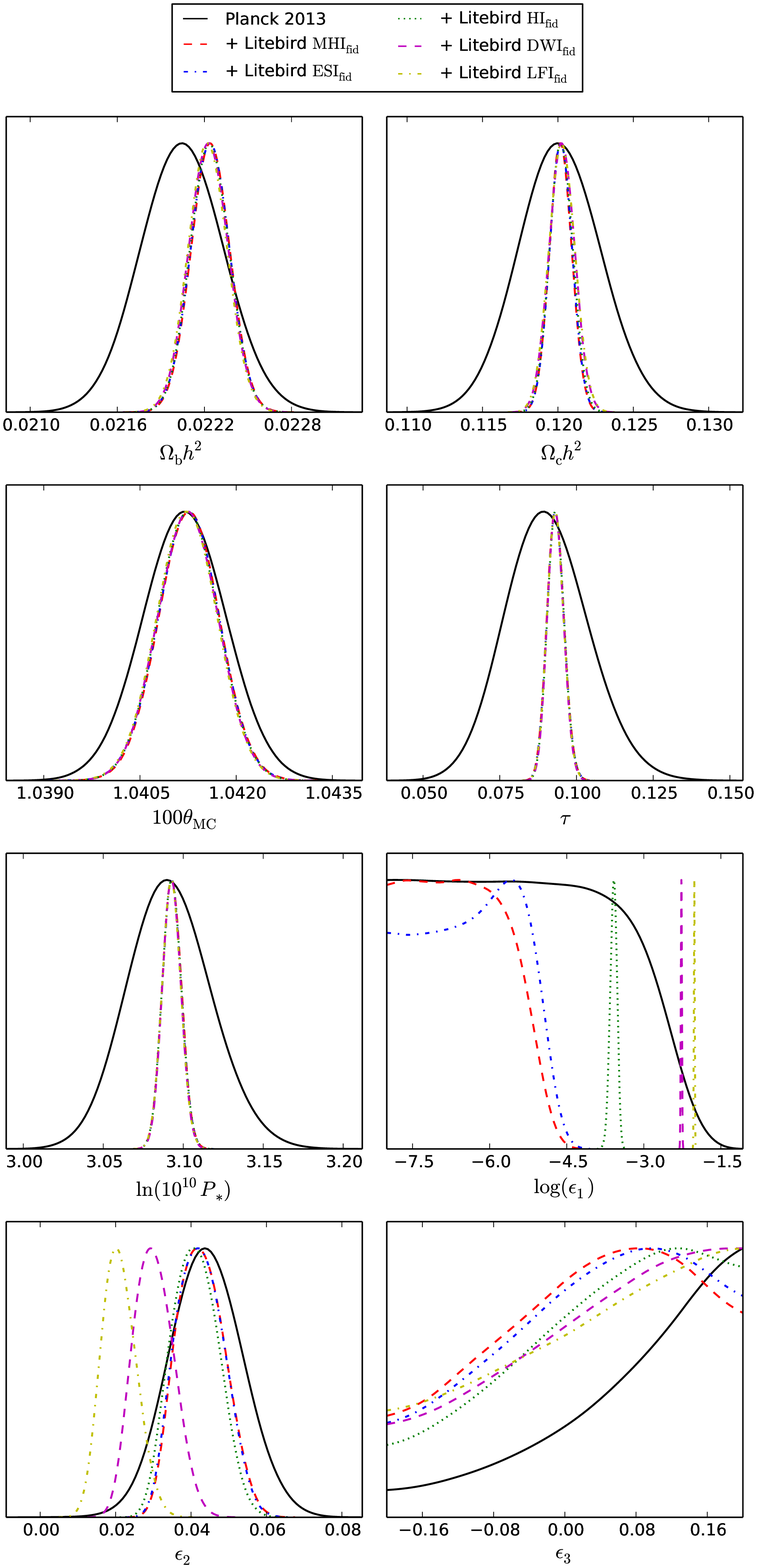}
\end{center}
\caption{PRISM (left) and LiteBird+Planck (right) one-dimensional marginalized
  posterior distributions of the post-inflationary ($\OmegaB h^2$,
  $\OmegaCDM h^2$, $100\thetaMC$, $\tau$) and primordial ($\Pstar$,
  $\epsstar{1}$, $\epsstar{2}$, $\epsstar{3}$) parameters for the five
  fiducial models, compared to the Planck 2013 posteriors (see
  legend).}
\label{fig:sr2ndlog}
\end{figure}


We start this section by presenting the one-dimensional marginalized
distributions of the $\Lambda$CDM and inflationary parameters for
PRISM and LiteBIRD+Planck in \Fig{fig:sr2ndlog}. Concerning the
post-inflationary parameters, and $\Pstar$, the corresponding
posteriors are strongly peaked at their fiducial values and this is
compatible with previous results for PRISM~\cite{Andre:2013nfa}. For
LiteBIRD+Planck, it is instructive to compare these posteriors to the
ones of Planck 2013 alone (also represented in \Fig{fig:sr2ndlog}) as
any improvement necessarily comes for LiteBIRD alone.

More interestingly, we notice that for all the fiducial models
considered, the posteriors for the cosmological parameters remain the
same showing that they do not correlate significantly with the
primordial parameters.

Regarding the slow-roll parameters, $\epsilon_1$ is, as expected, well
determined for $\lfiFID$, $\dwiFID$ and $\hiFID$. For $\esiFID$ and
$\mhiFID$, the amount of primordial tensor mode is under the detection
threshold and only an upper bound on $\epsilon_1$ can be
extracted. Let us remark the small bump for $\esiFID$ which is at
$\epsilon_1\simeq 3\times 10^{-6}$ while the underlying fiducial value
is $\epsilon_1 \equiv 3.28 \times 10^{-6}$. The second slow-roll
parameter $\epsilon_2$ is also well inferred and always peaked at its
fiducial value which is not surprising as it encodes the tilt of the
scalar power spectrum. One remarks that the PRISM precision on this
parameter is much better than the LiteBIRD+Planck one.

The situation concerning the third slow-roll parameter $\epsilon_3$,
which appears at second order in slow roll and determines the
running of the power spectra, is completely new~\cite{inprep}. We see
that PRISM would be able to constrain this parameter thereby adding a
new observable for the inflationary dynamics. This is not the case for
LiteBIRD+Planck and can be traced to the range of angular scales
accessible to PRISM due to both its resolution and sensitivity.

\begin{figure}
\begin{center}
\includegraphics[width=\wsingfig]{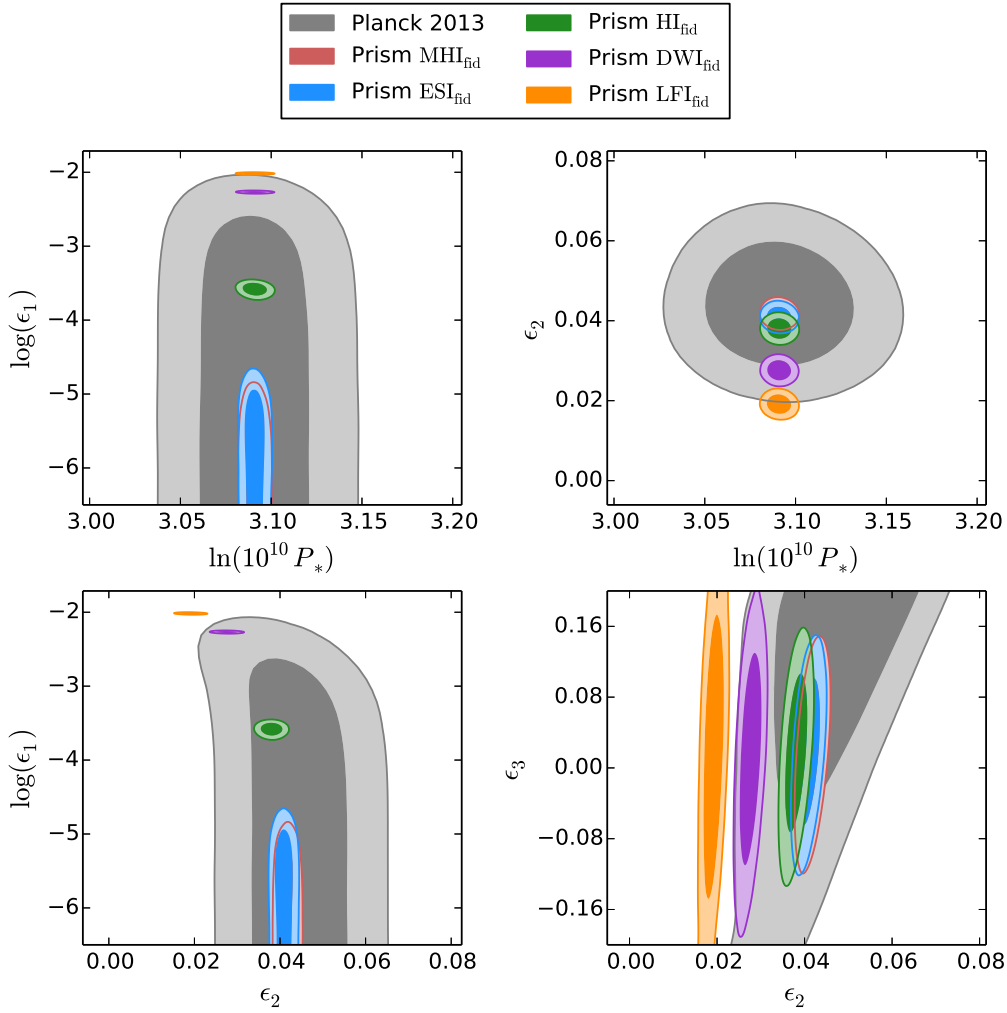}
\end{center}
\caption{PRISM two-dimensional marginalized posterior distributions of
  the slow-roll parameters ($\Pstar$, $\epsstar{1}$, $\epsstar{2}$,
  $\epsstar{3}$) for the five fiducial models considered, compared to
  the Planck 2013 two-dimensional posterior distributions (black
  shaded region). The $\mhiFID$ red shaded region is almost entirely
  behind the $\esiFID$ blue shaded region.}
\label{fig:prism_sr2ndlog_2D}
\end{figure}

\begin{figure}
\begin{center}
\includegraphics[width=\wsingfig]{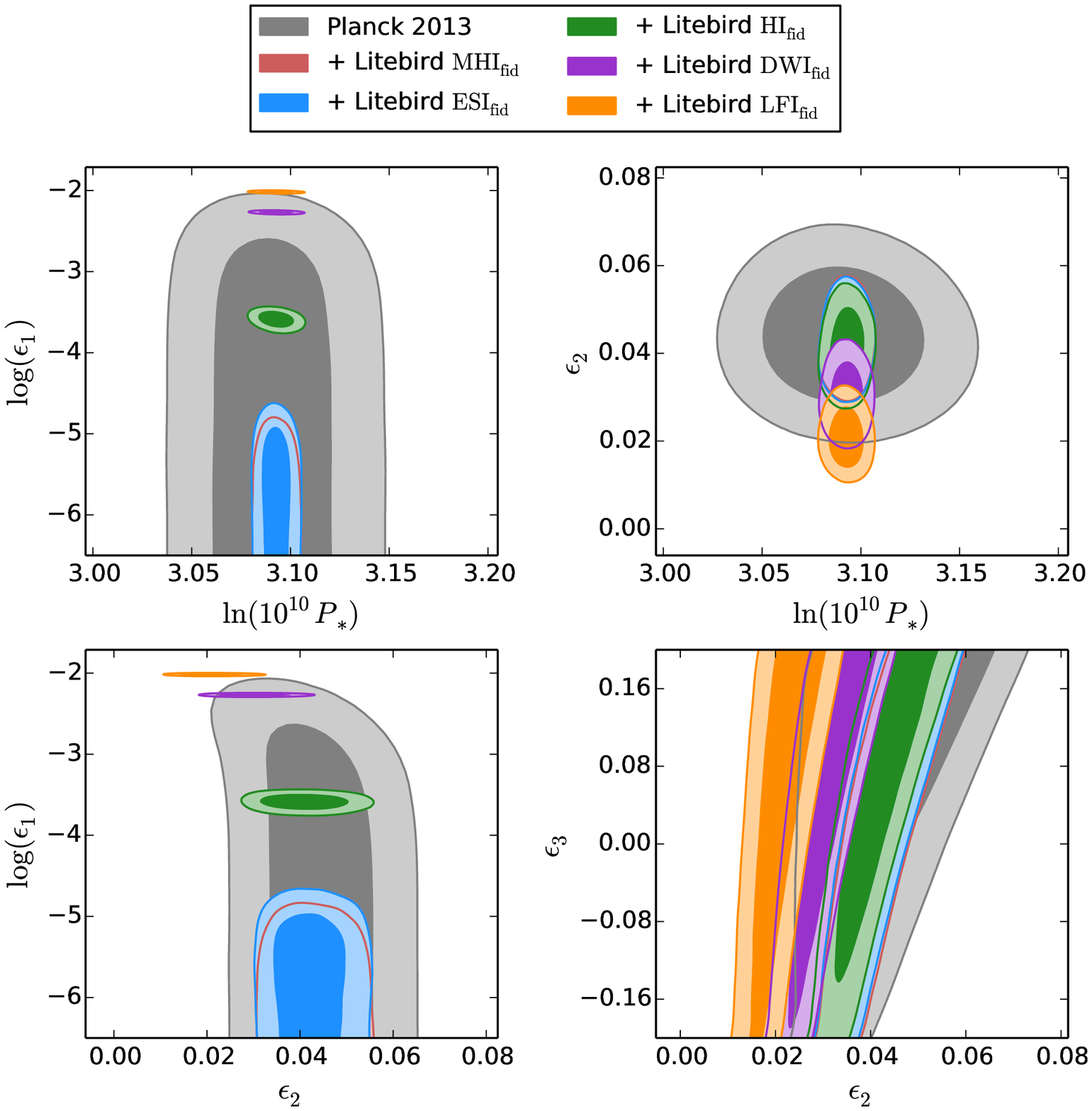}
\end{center}
\caption{LiteBIRD+Planck two-dimensional marginalized posterior distributions
  of the slow-roll parameters ($\Pstar$, $\epsstar{1}$, $\epsstar{2}$,
  $\epsstar{3}$) for the five fiducial models considered, compared to
  the Planck 2013 two-dimensional posterior distributions (black
  shaded region). The $\mhiFID$ red shaded region is almost entirely
  behind the $\esiFID$ blue shaded region.}
\label{fig:litebird_sr2ndlog_2D}
\end{figure}

In Figs.~\ref{fig:prism_sr2ndlog_2D} and
\ref{fig:litebird_sr2ndlog_2D}, we present the two-dimensional
marginalized probability distributions for the primordial parameters
$\Pstar$, $\epsilon_1$, $\epsilon_2$, and $\epsilon_3$. The
two-dimensional posteriors in the plane $\left(\log
\epsilon_1,\epsilon_2\right)$ illustrate the differences in design
between PRISM and LiteBIRD. The sensitivity of both experiments in the
$B$-modes gives very strong constraints on $\epsilon_1$ when the
fiducial model lies above the detection threshold. However, as opposed
to PRISM, the low angular resolution of LiteBIRD does not allow to
significantly improve the determination of $\epsilon_2$ compared to
Planck alone. Finally, let us again notice that PRISM yields closed
contours for the two-sigma confidence intervals in the plane
$(\epsilon_2,\epsilon_3)$.

\begin{figure}
\begin{center}
\includegraphics[width=\wdblefig]{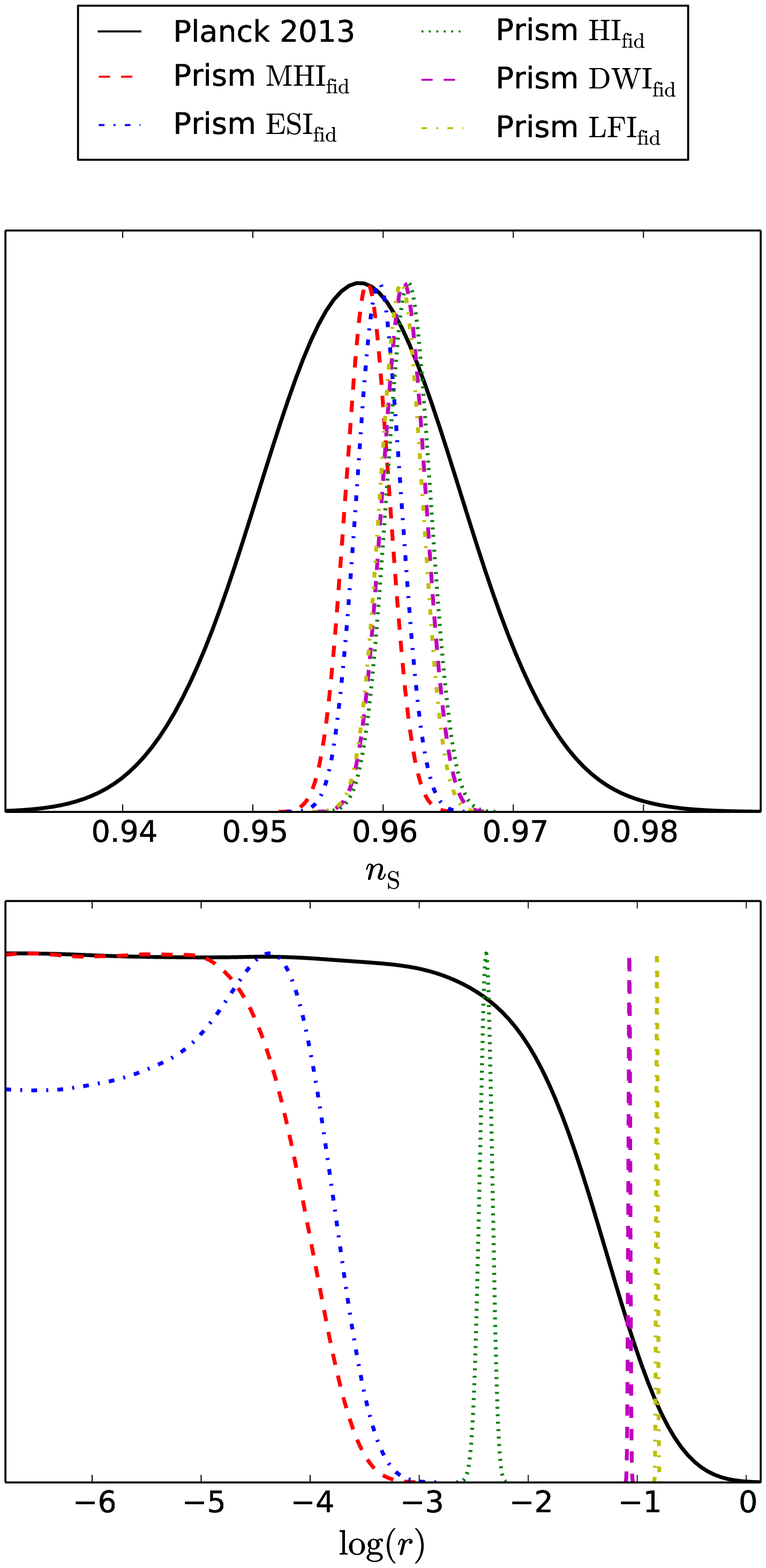}
\includegraphics[width=\wdblefig]{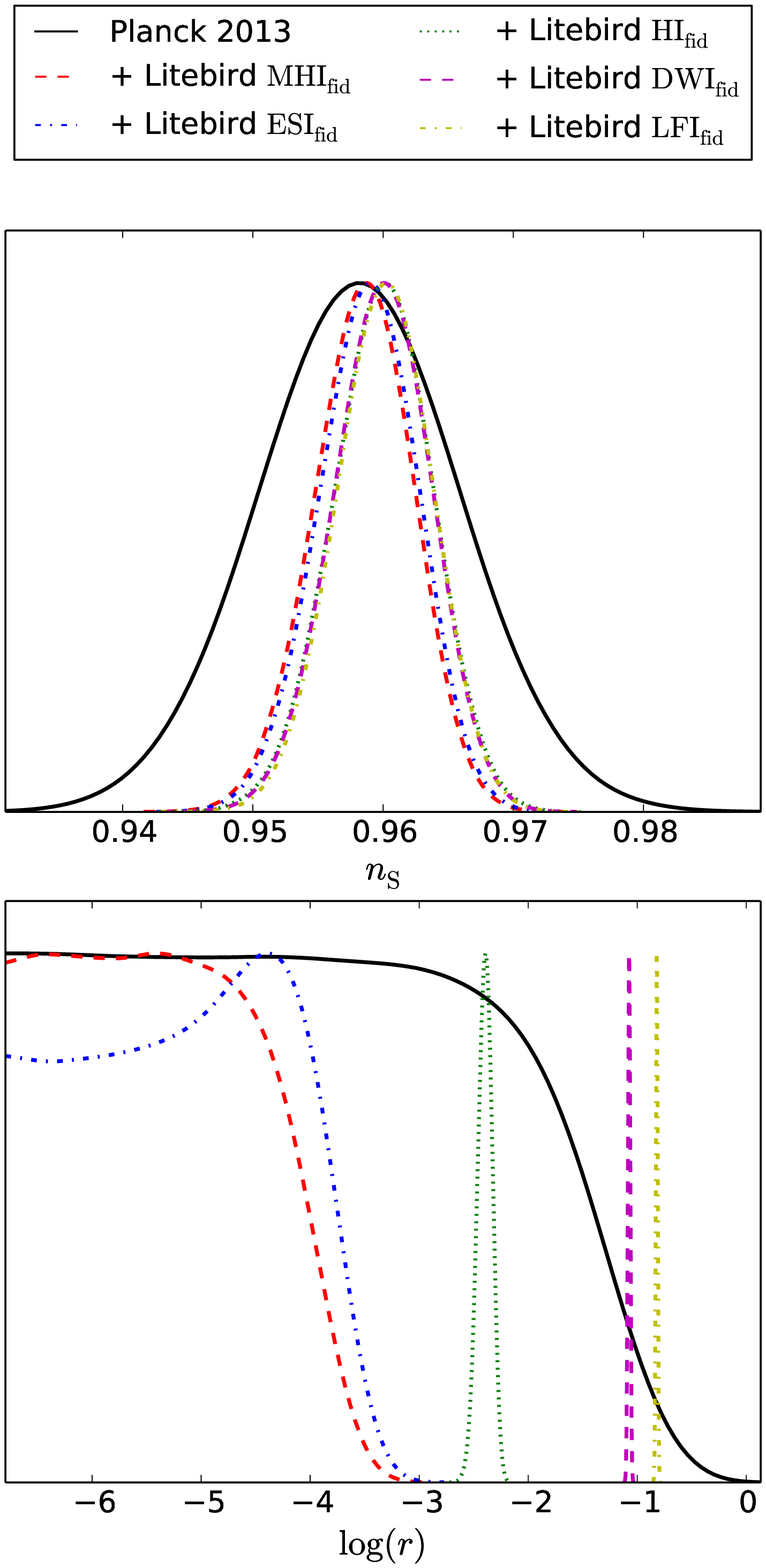}
\end{center}
\caption{PRISM (left) and LiteBIRD+Planck (right) one-dimensional
  marginalized posterior distributions of the scalar spectral index
  $\nS$ (top panels) and of the tensor-to-scalar ratio $r$ (bottom
  panels) for the same five fiducial models compared to the Planck
  posterior distributions (solid black line). They have been obtained
  from importance sampling based on the slow-roll parameter posteriors
  (see Figs.~\ref{fig:prism_sr2ndlog_2D} and
  \ref{fig:litebird_sr2ndlog_2D}).}
\label{fig:pllog}
\end{figure}

\begin{figure}
\begin{center}
\includegraphics[width=\wdblefig]{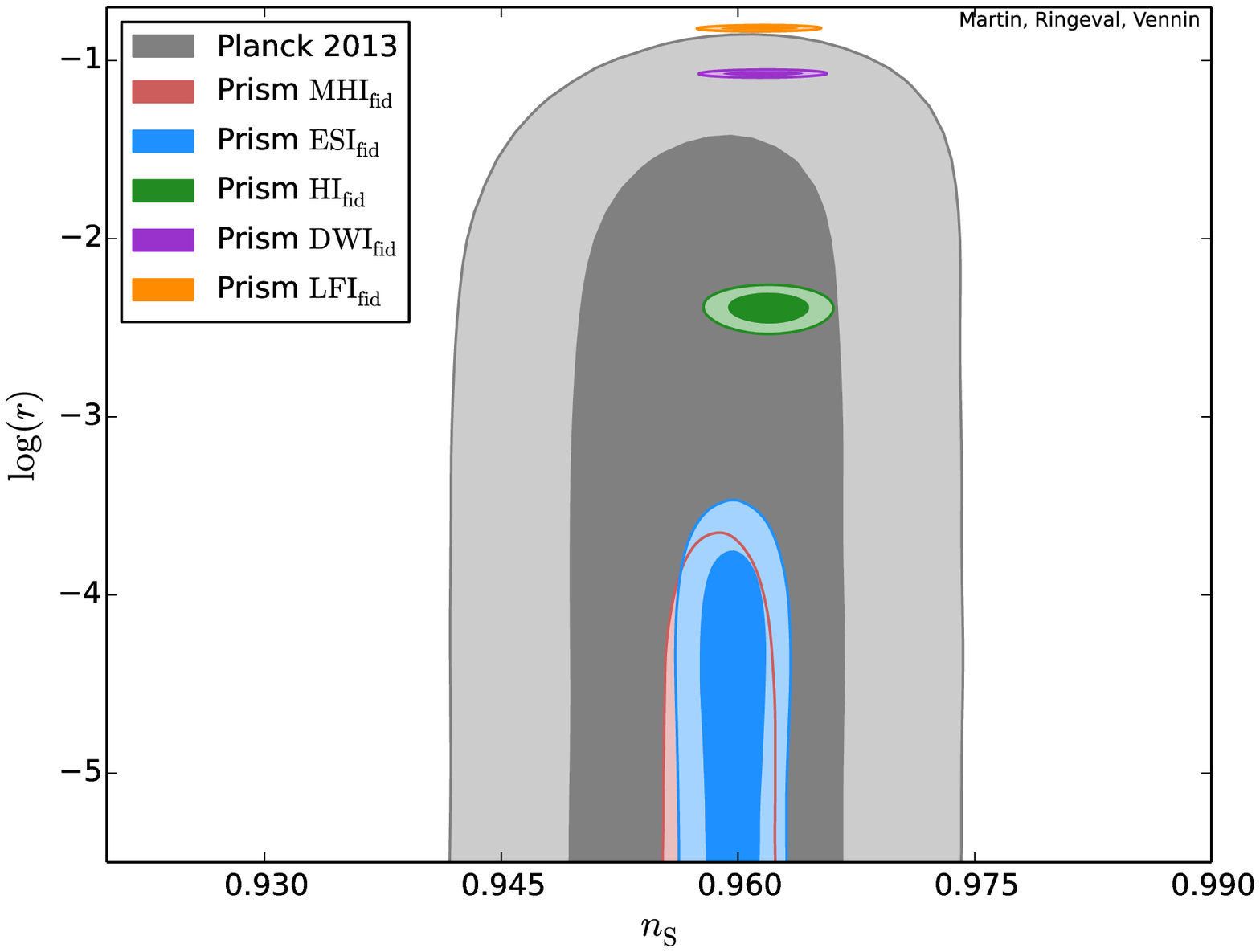}
\includegraphics[width=\wdblefig]{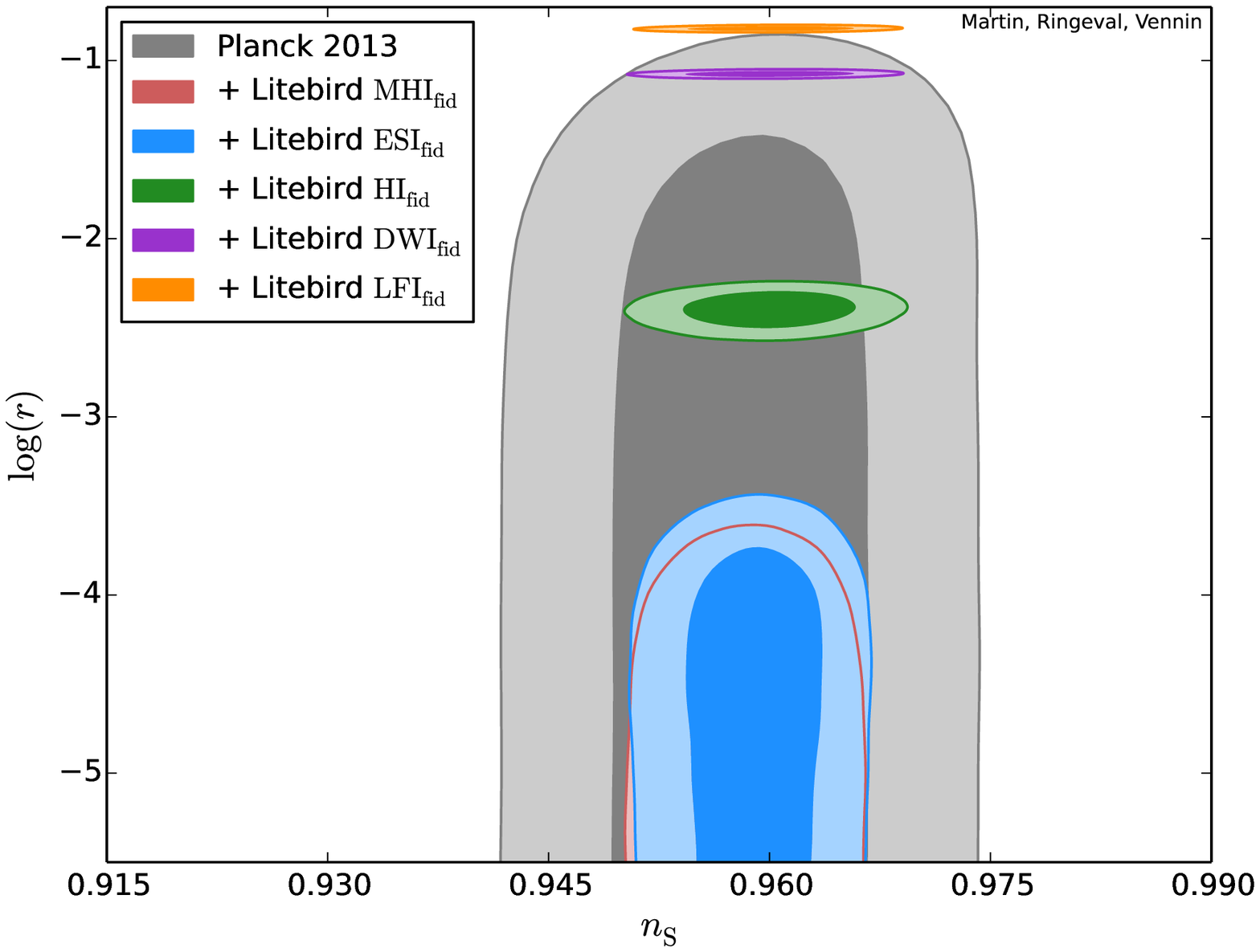}
\end{center}
\caption{PRISM (left) and LiteBIRD (right) two-dimensional
  marginalized posterior distributions of the derived power law
  parameters $(\nS, r)$ for the five fiducial models under scrutiny
  compared to the Planck two-dimensional posterior distributions
  (black shaded region). The $\mhiFID$ red shaded region is almost
  entirely hidden behind the one of $\esiFID$.}
\label{fig:nslogr_2D}
\end{figure}

By importance sampling, one can also infer the so-called power-law
parameters. In particular, the scalar spectral index $\nS$ and the
tensor-to-scalar ratio $r$ are analytic functions of the slow-roll
parameters. At second order in slow roll, $\nS$ is given by
$\nS=1-2\epsilon_1 - \epsilon_2 - 2\epsilon_1^2 - (2C+3)\epsilon_1
\epsilon_2 -C \epsilon_2 \epsilon_3$ and $r= 16\epsilon_1 + 16C
\epsilon_1 \epsilon_2$ (the parameter $C \simeq -0.73$ is a numerical
constant). The corresponding one-dimensional marginalized
distributions for $\nS$ and $r$ are represented in \Fig{fig:pllog} for
PRISM and LiteBIRD+Planck. One notices that, for the five fiducial
models, the scalar spectral index is well reconstructed and that its
posterior distribution peaks at its fiducial value. For $\lfiFID$,
$\dwiFID$ and $\hiFID$, this is because the posterior distributions of
$\epsilon_1$ and $\epsilon_2$ are also well constrained. For $\esiFID$
and $\mhiFID$, there is only an upper bound on $\epsilon_1$ but, since
it is a very small parameter compared to $\epsilon_2$, one has in fact
$\nS \simeq 1-\epsilon_2$. Therefore, up to the second order
corrections, the posterior of $\nS$ is essentially driven by the one
on $\epsilon_2$. In particular, the difference between the width of
these posteriors also comes from different beam resolution between
LiteBIRD and PRISM.

Concerning the posterior of $r$, the discussion is essentially similar
to the one about the posterior distribution of $\epsilon_1$ (since
these two parameters are proportional at leading order in
slow roll). The quantity $r$ is strongly peaked at its fiducial value
for $\lfiFID$, $\dwiFID$ and $\hiFID$, while it is only constrained
from above for $\esiFID$ and $\mhiFID$.

Finally, in \Fig{fig:nslogr_2D}, we have represented the
two-dimensional posterior distribution in the plane $(\nS, \log r)$ for the
five different cases studied in this paper.

\begin{figure*}
\begin{center}
\includegraphics[width=\wsingfig]{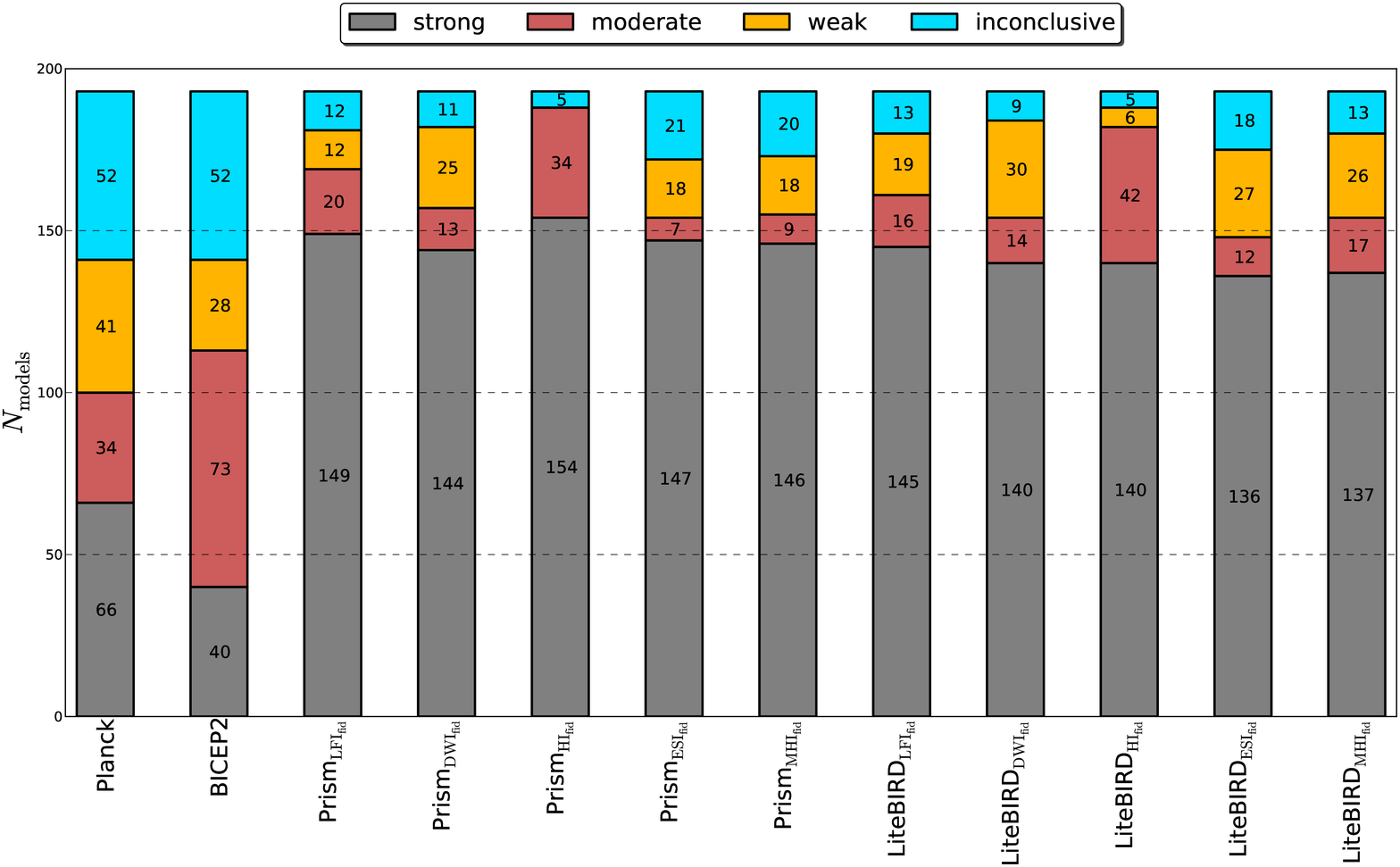}
\end{center}
\caption{Distribution of the {\EI} models within each Jeffreys'
  categories for PRISM and LiteBIRD+Planck and for the five different
  fiducial models. The Planck~\cite{Martin:2013nzq} and
  BICEP2~\cite{Martin:2014lra} results have been reported for the sake
  of comparison.}
\label{fig:histo}
\end{figure*}

Let us now turn to the computation of the Bayesian evidences. In
Fig.~\ref{fig:histo}, for both PRISM and LiteBIRD as well as for each
fiducial models ($\lfiFID$, $\dwiFID$, $\hiFID$, $\esiFID$ and
$\mhiFID$), we have reported the number of {\EI} models in each of the
four Jeffreys' categories giving the strength of belief that the model
is explaining the data compared to the best. We have also reported the
results obtained in \Refc{Martin:2013nzq} for Planck 2013 and in
\Refc{Martin:2014lra} for BICEP2.

One notices the remarkable constraining power of these two
experiments: PRISM is able to rule out (``strong'' in the Jeffreys'
scale) more than three quarters of the inflationary models in all the
situations we have studied, $77\%$ if $\lfiFID$ is the fiducial model,
$75\%$ for $\dwiFID$, $80\%$ for $\hiFID$, $76\%$ for $\esiFID$ and
$76\%$ for $\mhiFID$. These numbers are similar (although slightly
smaller) for LiteBIRD+Planck, namely $75\%$ for $\lfiFID$, $73\%$ for
$\dwiFID$, $73\%$ for $\hiFID$, $70\%$ for $\esiFID$ and $71\%$ for
$\mhiFID$. One should compare these performances to the Planck (and
BICEP2) results that are able to respectively reject $34\%$ and $21\%$
of the inflationary scenarios. It is also interesting to remark that
the percentage of models in the strong zone does not depend a lot on
the assumed fiducial model and, therefore, appears to be quite
generic.

Let us now discuss the opposite end of the Jeffreys' scale, namely the
models in the ``inconclusive'' zone.

For PRISM with $\lfiFID$ as fiducial model, we find that only $6\%$ of
the scenarios are in this category. If, in addition, we take into
account the Bayesian complexity and restrict ourselves to models
having a number of unconstrained parameter between zero and one, then
this number falls to $4\%$.

In the case of the fiducial model $\dwiFID$, one finds very similar
numbers. For PRISM, we get $6\%$ in the ``inconclusive'' zone and
$3\%$ in the ``inconclusive'' zone having a number of unconstrained
parameters between zero and one.

If the fiducial model is now the Starobinsky model, $\hiFID$, then the
performances are even better: only $3\%$ of the models are in the
inconclusive zone and, if one considers only the scenarios with a
number of unconstrained parameters between zero and one, then one
singles out a subset of three models only, corresponding to $1.5\%$ of
the total number of {\EI} scenarios. It appears that if the
inflationary model actually realized in Nature is similar to $\hiFID$,
then PRISM and LiteBIRD+Planck would provide an optimal setting.

In the case where $\esiFID$ is the fiducial model, $11\%$ of the
scenarios are in the inconclusive zone (and $4\%$ if the number of
unconstrained parameters is taken between zero and one). Let us recall
that, in this situation, and contrary to the three preceding examples,
the detection of primordial gravity waves is no longer
possible.

Finally, the case where $\mhiFID$ is the fiducial scenario is quite
similar. As for $\esiFID$, this corresponds to a situation where the
tensor-to-scalar ratio $r$ is too small to be measured, even by the
PRISM experiment. One finds $11\%$ in the inconclusive category and
$4\%$ if Bayesian complexity is also taken into account.

For LiteBIRD+Planck, these figures are almost the same while being in
a non-significant way slightly less constraining (see
\Fig{fig:histo}). Only for $\esiFID$ and $\mhiFID$, one finds less
models in the inconclusive zone ($9\%$ and $7\%$, respectively), but
this is due to a boundary effect as there are more models in the weak
evidence zone.

In view of these results, a word of caution is in order. Indeed, one
might naively conclude that a simple design like LiteBIRD, combined
with Planck, would perform as well as a big mission like
PRISM. However, here, we have assumed perfect foregrounds removal for
both LiteBIRD and PRISM. This may be justified for PRISM because the
mission is precisely designed to accurately measure various other
astrophysical signals (such as the Cosmic Infrared Background, etc...)
in addition to the CMB thereby allowing an optimal components
separation~\cite{Andre:2013nfa}. This might not be the case for
LiteBIRD and our forecasts for LiteBIRD+Planck may be
over-idealized. Moreover, as it should be clear from
\Fig{fig:sr2ndlog}, only PRISM will have the ability to measure
$\epsilon_3$, that is to say the slow-roll running of the scalar power
spectrum~\cite{inprep}. This is not so relevant for the model
comparison here, precisely because all slow-roll models necessarily
produce small values of $\epsilon_3$. However, if slow-roll ends up
being violated at second order ($\epsilon_3>1$), PRISM will have the
ability to detect it.

\section{Conclusions}
\label{sec:conclusion}

Let us now summarize our main findings. In this paper, we have studied
the ability to constrain the inflationary theory of two post-Planck
CMB missions: PRISM and LiteBIRD. Our method consists in simulating
CMB data for five different inflationary scenarios corresponding to a
representative sample of inflationary models, leading to different
values of the tensor-to-scalar ratio $r$ smaller and larger that the
target of those mission $r\simeq 10^{-3}$. For each mock data, we have
computed the Bayesian evidences and complexities of all {\EI} models
and have studied how they are distributed among the four Jeffreys'
categories with respect to the best model. We have found that the
number of models that can be ruled out at a statistically significant
level typically goes from one third for Planck to three quarters for
PRISM and LiteBIRD+Planck. The gain in constraining power is therefore
significant, illustrating the efficiency of constraining the
observable $r$ (or $\epsilon_1$) as well as improving the measurement
accuracy on the other primordial parameters.

Let us stress again that our results have been derived in what we have
called ``the worst case scenario'', namely for single-field slow-roll
models. Indeed, these are considered to be the most difficult to infer
as they produce an undetectable amount of non-Gaussianities and do not
generate entropy perturbations. In a wider framework in which one
would consider non-minimal inflationary models, one could only expect
stronger constraints (see for instance \Refcs{Andre:2013nfa,
  Clesse:2014pna}). In this context, we have found that PRISM could
rule out slow-roll inflation by its ability to measure a less than
unity second order slow-roll parameter $\epsilon_3$, a result which
has been discussed before only in the context of future 21cm
experiments~\cite{Adshead:2010mc, Clesse:2012th}.

Finally, let us discuss possible improvements of the present work.
Going further than estimating the constraining power of the future CMB
missions, it would be interesting to investigate the model
identification problem. For instance, for the PRISM mission, we have
checked that the Bayesian evidence of $\dwiTFIVE$, a model sharing
exactly the same potential shape as $\dwiFID$ but having an unknown
reheating temperature and unknown potential normalization ($\Pstar$)
would end up being moderately disfavored compared to
$\dwiFID$. Another model $\dwiRTFIVE$, taken to be of same potential
shape and reheating temperature than $\dwiFID$, but still having an
unknown potential normalization, would remain within the inconclusive
region compared to $\dwiFID$. This suggests that identifying the
``correct'' inflationary scenario with a mission like PRISM will
necessitate to have a good prior knowledge on the reheating energy
scale. Conversely, this also suggests that blindly looking at the
potential shape without specifying how the reheating proceeds could
lead to false positive identifications of the inflationary model. In
other words, these future CMB missions will reach an accuracy such
that specifying couplings of the inflaton to the Standard Model of
particle physics will become compulsory~\cite{GarciaBellido:2008ab}
and will actually be seen in the CMB sky~\cite{inprep}.

\acknowledgments

This work is partially supported by the ESA Belgian Federal PRODEX
Grant No.~4000103071 and the Wallonia-Brussels Federation grant ARC
No.~11/15-040. Some figures have been made owing to the
{\GetDistPlots} python scripts provided with the {\COSMOMC}
code~\cite{Lewis:2002ah}.

\bibliography{biblio}
\bibliographystyle{JHEP}

\end{document}